\documentclass{aip-cp}

\usepackage[numbers]{natbib}
\usepackage{rotating}
\usepackage{graphicx}
\usepackage{mathrsfs}
\usepackage{amssymb}
\usepackage{graphicx}
\usepackage[normalem]{ulem}
\usepackage{bm}
\usepackage{longtable}
\usepackage{times}
\usepackage{subfigure}

\renewcommand\sout{\bgroup \color{red} \ULdepth=-.5ex \ULset}

\renewcommand{\rm}[1]{\textrm{#1}}

\newcommand{\pt}{\partial}

\begin{document}

\title{Dense matter equation of state and neutron star properties from nuclear theory and experiment}
\author[aff1,aff2]{Jeremy W. Holt\corref{cor1}}
\author[aff1,aff2]{Yeunhwan Lim}
\corresp[cor1]{Corresponding author and speaker: holt@physics.tamu.edu}
\affil[aff1]{Department of Physics and Astronomy, Texas A\&M University, College Station,
Texas, 77840, USA}
\affil[aff2]{Cyclotron Institute, Texas A\&M University, College Station, TX 77840, USA}
\maketitle

\begin{abstract}
The equation of state of dense matter determines the structure of neutron stars, their typical radii, and maximum masses. Recent improvements in theoretical modeling of nuclear forces from the low-energy effective field theory of QCD has led to tighter constraints on the equation of state of neutron-rich matter at and somewhat above the densities of atomic nuclei, while the equation of state and composition of matter at high densities remains largely uncertain and open to a multitude of theoretical speculations. In the present work we review the latest advances in microscopic modeling of the nuclear equation of state and demonstrate how to consistently include also empirical nuclear data into a Bayesian posterior probability distribution for the model parameters. Derived bulk neutron star properties such as radii, moments of inertia, and tidal deformabilities are computed, and we discuss as well the limitations of our modeling.
\end{abstract}

\section{Introduction}
\label{intro}
The fate of matter at ultra-high densities ($\rho \simeq 10^{15}$\,g/cm$^3$) found in the compressed inner cores of neutron stars is one of the fundamental science questions to be investigated through upcoming observational campaigns of neutron stars. The crust and outer core of neutron stars are governed by properties of the strong nuclear force that are constrained by nucleon-nucleon scattering and bound state data of atomic nuclei (binding energies, radii, and collective excitations). However, as the density or proton-neutron asymmetry increases, uncertainties in nuclear two-body and especially three-body forces become increasingly important, leading to the growth of theoretical errors in the equation of state. Since the bulk properties of neutron stars, such as radii, moments of inertia, and tidal deformabilities, are strongly correlated \cite{lattimer01,lim18a,tsang19} with the pressure of beta-equilibrium nuclear matter at about twice the density of atomic nuclei, it is expected that upcoming neutron star observations will place novel constraints on the properties of dense matter beyond the reach of present nuclear theory modeling. What remains to be seen is whether these observational campaigns will identify the imprints of exotic matter or phase transitions that may occur in neutron star inner cores at several times normal nuclear densities.

Already the first observation of gravitational waves from the binary neutron star merger event GW170817 has been shown to favor relatively soft equations of state, leading to upper bounds on the radii of neutron stars, tidal deformabilities, and moments of inertia \cite{lim18a,tsang19,Fattoyev18,annala18,most18,krastev18,tews2018gw,landry18,abbott2018b,lim18b}. In addition, upper bounds on the maximum neutron star mass have been inferred from combined analysis of the electromagnetic counterpart (kilonova) and numerical simulations of neutron star mergers \cite{bauswein17,margalit17,shibata17,radice18,rezzolla18,ruiz18}. The LIGO/VIRGO third observing run in 2019 is positioned to add multiple detections of neutron star mergers, enabling refined constraints on the nuclear equation of state and bulk neutron star properties. Moreover, the NICER X-ray telescope will provide complementary constraints on the mass-radius relationship of neutron stars, with estimated neutron star radius measurements at the 5\% precision level. A high priority in nuclear theory is the development of models capable of relating these astronomical observations to properties of nuclei, nuclear matter, and the nuclear force.

\subsection{Equation of State Parametrization}
\label{eos}
The dense matter equation of state at and somewhat above normal nuclear densities can be inferred from the properties of heavy nuclei \cite{dutra12,lattimer13}, data from medium-energy heavy-ion collisions \cite{danielewicz02,shetty07,qin12}, microscopic calculations based on high-precision two- and three-body forces \cite{hebeler10,gandolfi2012,gezerlis13,roggero14,wlazlowski14,drischler14,sammarruca15,gandolfi15,holt17prc}, and neutron star observations \cite{steiner10,raithel16,raithel17}. For soft equations of state favored by GW170817, the central densities in neutron stars range from about $3-7n_0$, where $n_0 = 0.16$\,fm$^{-3}$ is the saturated central density in heavy nuclei. A central question is how well the equation of state constrained by nuclear theory and experiment can be extrapolated to larger densities in neutron stars. From explicit calculations \cite{holt18} of the energy per particle of pure neutron matter $\bar E_{pnm}(n)$ and symmetric nuclear matter $\bar E_{snm}(n)$ (as well as their difference at a specific number density $n$, which defines the symmetry energy $E_{sym}(n) = \bar E_{pnm}(n) - \bar E_{snm}(n)$), it is found that a Taylor series expansion in powers of the Fermi momentum converges well at nuclear saturation density when the expansion point is taken to be near $n_0/2$. At the reference density $n_0/2$, nuclear experiments and theory give fairly tight constraints \cite{brown13,lynch18} on the symmetry energy and therefore precise extrapolations to nuclear saturation density. Below $n_0/2$, logarithmic contributions \cite{kaiser02} to the symmetry energy strongly reduce the radius of convergence, thereby providing a lower bound on the extrapolation point \cite{holt18}. A  Taylor series expansion of the symmetry energy around saturation density
\begin{equation}
E_{sym} = J + L \left(\frac{n-n_0}{3n_0} \right) + \frac{1}{2}K_{\mathrm{sym}}\left(\frac{n-n_0}{3n_0} \right)^2
   + \frac{1}{6}Q_{\mathrm{sym}}\left(\frac{n-n_0}{3n_0} \right)^3 + \cdots
\label{symtay}
\end{equation}
is commonly employed in the literature, and efforts to constrain the empirical parameters $J, L, K_{sym},$ and $Q_{sym}$ (as well as their correlations) may provide important insights into the properties of dense matter beyond nuclear saturation density.

In recent years, chiral effective field theory \cite{weinberg79} has developed into a powerful tool for analyzing low-energy hadronic systems and in particular for deriving nuclear forces \cite{epelbaum09,machleidt11}. Chiral effective field theory is expected to be well converged for systems where the characteristic momentum scale is below the chiral symmetry breaking scale $\Lambda_\chi$ $\simeq 1$\,GeV. However, in practice chiral nuclear potentials contain a momentum-space regulating function with cutoff parameter $\Lambda \simeq 500-600$\,MeV that is well below $\Lambda_\chi$. This limits the application of chiral effective field theory to densities not far above that of normal nuclei, since the Fermi momenta for symmetric nuclear matter and pure neutron matter at that density have the values $k_F \simeq 260$\,MeV and $k_F \simeq 330$\,MeV, respectively. In the case of many-body perturbation theory calculations of the equation of state, theoretical uncertainties can be estimated by varying (i) the order in the chiral expansion, (ii) the value of the momentum-space cutoff $\Lambda$, and (iii) the order in many-body perturbation theory. In previous calculations \cite{sammarruca15, holt17prc}, these variations led to only moderate uncertainties up to $n=2n_0$, which is surprising since the estimated breakdown scale is well below $2n_0$. Additional variations, such as the choice of observables used to fit the three-body contact terms as well as the form of the regulating function (local vs.\ nonlocal) \cite{dyhdalo16} should therefore be explored in future work.

\begin{figure}[t]
\centering
\includegraphics[scale=0.75]{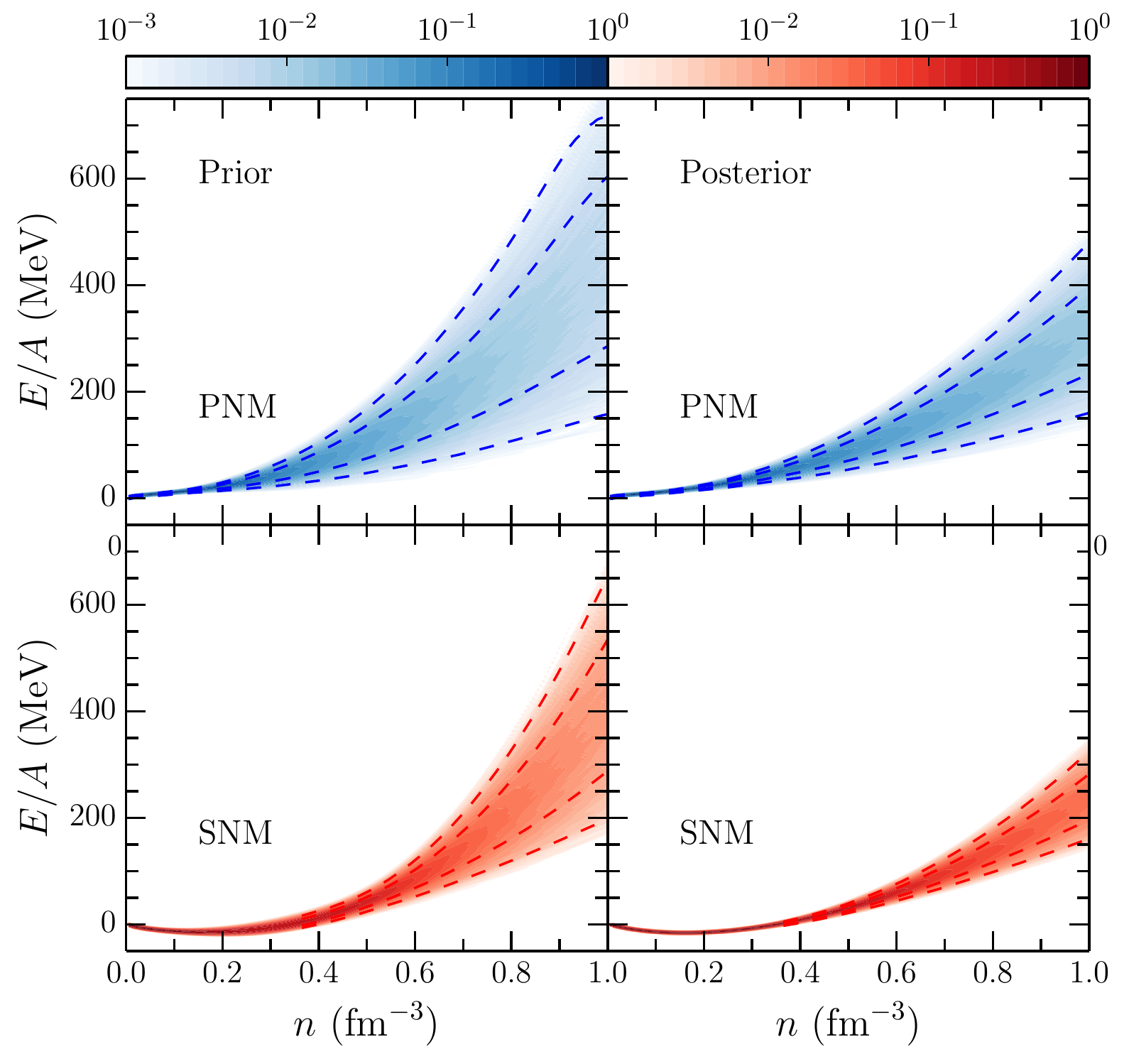}
\caption{Prior and posterior probability distributions for the zero-temperature equation of state for pure neutron matter (PNM) and symmetric nuclear matter (SNM) obtained from the Bayesian analysis in the present work. The dashed lines indicate the $1\sigma$ and $2\sigma$ probability contours.}
\label{prior}
\end{figure}

In the present set of chiral effective field theory calculations of the nuclear equation of state, we vary the order in the chiral expansion from next-to-next-to-leading order (N2LO) to N3LO, the cutoff $\Lambda$ from $400 - 500$\,MeV, and the order in many-body perturbation theory from second to third order (also including second-order self-energy corrections in the latter case). In previous works \cite{lim18a,lim18b,lim19} we have parametrized the resulting equation of state through a set of energy density functionals of the form
\begin{equation}
\label{eq:edf}
\mathcal{E}(n,x) =  \frac{1}{2m}\tau_n + \frac{1}{2m}\tau_p + [1-(1-2x)^2] f_s(n) + (1-2x)^2 f_n(n) \,,
\end{equation}
where $\tau_p$ ($\tau_n$) is the kinetic energy density of protons (neutrons), $x = n_p/(n_n + n_p)$ is the proton fraction, and $f_s$ ($f_n$) is the potential energy density for symmetric nuclear matter (neutron matter) expanded in powers of the Fermi momentum:
\begin{equation}
\label{eq:fns}
f_s(n) = \sum_{i=0}^{3} A_i\, n^{(2+i/3)} \,, 
\quad
f_n(n) = \sum_{i=0}^{3} B_i\,n^{(2+i/3)}\,.
\end{equation}
For definiteness we write the above expressions as expansions about a reference Fermi momentum $k_F^r$:
\begin{equation}
\frac{\mathcal{E}}{n}(k_F,x=0.5) = 2^{2/3}\frac{3}{5}\frac{k_F^2}{2m} + \frac{k_F^3}{9\pi^2} \sum_{i=0}^{3} \frac{a_i}{i!}\, \beta^i
\label{eq:exp1}
\end{equation}
\begin{equation}
\frac{\mathcal{E}}{n}(k_F,x=0) = \frac{3}{5} \frac{k_F^2}{2m} + \frac{k_F^3}{9\pi^2} \sum_{i=0}^{3} \frac{b_i}{i!}\, \beta^i\,,
\label{eq:exp2}
\end{equation}
where $\beta = (k_F - k_F^r) / k_F^r$. In both Eqs.\ (\ref{eq:exp1}) and (\ref{eq:exp2}) we define $k_F = ( 3 \pi ^ 2 n ) ^ {1/3}$. Fitting the equations of state from chiral effective field theory up to $n=0.32$\,fm$^{-3}$ to the form given in Eqs.\ (\ref{eq:exp1}) and (\ref{eq:exp2}) yields a joint probability distribution for the $a_i$ and $b_i$ parameters. The resulting distributions for the equation of state are shown in the left columns of Fig.\ \ref{prior} and labeled ``Prior''.  Later we will lower the fitting range to $n<0.25$\,fm$^{-3}$ to explore the associated uncertainties.

An advantage of microscopic modeling is that the underlying force is fitted only to the properties of two- and three-nucleon systems. Therefore, the equation of state is essentially calculated independently of the properties of medium-mass and heavy nuclei, which provide further experimental constraints on the equation of state around nuclear matter saturation density and for nearly isospin-symmetric systems. To exploit this feature, we have developed \cite{lim18a,lim18b,lim19} a Bayesian approach in which microscopic calculations of the nuclear matter ground state energy determine the prior distribution functions for the model parameters in Eqs.\ (\ref{eq:exp1}) and (\ref{eq:exp2}). Information about heavy nuclei binding energies, charge radii, and collective excitation modes are then built into Bayesian likelihood functions (see also Ref.\ \cite{margueron18}). Modern nuclear energy density functionals, such as Skyrme and Relativistic Mean Field models, are fitted to this wide range of nuclear data, and therefore we take as a proxy for experimental data a large set of 205 Skyrme energy density functionals \cite{dutra12} from which we can extract Bayesian likelihood functions involving the $a_i$ parameters of Eq.\ (\ref{eq:exp1}). Since the properties of finite nuclei do not strongly constrain terms in the density-dependent symmetry energy beyond $J$ in Eq.\ (\ref{symtay}), we enforce correlations \cite{holt18,margueron19} between $J$, $L$, $K_{sym}$, and $Q_{sym}$ to obtain Bayesian likelihood functions involving the $b_i$ parameters in Eq.\ (\ref{eq:exp2}). We then construct posterior probability distributions according to Bayes' Theorem:
\begin{equation}
P(\mathbf{a}\vert \mathrm{data}) \sim P(\mathrm{data}\vert \mathbf{a})P(\mathbf{a}),
\label{bayes}
\end{equation}
where $P(\mathbf{a})$ is the prior distribution, $P(\mathrm{data}\vert \mathbf{a})$ is the likelihood function, and $P(\mathbf{a}\vert \mathrm{data})$ is the posterior distribution.
More specifically, the four parameters used to model the symmetric nuclear matter energy density are constrained by the nuclear saturation energy $B$, saturation density $n_0$, incompressibility $K$, and skewness $Q$ defined at $n=n_0$ and $x=1/2$:
\begin{equation}
B = - \frac{\mathcal{E}(n_0)}{n_0}\,, \quad
\left . p = n_0^2\frac{\partial (\mathcal{E}/n)}{\pt n}\right |_{n=n_0}=0\,, \quad \left . K = 9 n_0^2\frac{\partial^2 (\mathcal{E}/n)}{\pt n^2}\right |_{n=n_0}\,,\quad
\left . Q = 27n_0^3\frac{\partial^3 (\mathcal{E}/n)}{\pt n^3}\right |_{n=n_0}\,,
\end{equation}
where $p$ is the pressure at saturation density. In the case of pure neutron matter, we employ experimental constraints on the symmetry energy $J = 31 \pm 1.5$\,MeV \cite{lattimer13} together with correlations among $J$, $L$, $K_{\rm sym}$, and $Q_{\rm sym}$ \cite{holt18}.

\begin{figure*}[t]
\centering
\subfigure{\includegraphics[scale=0.44]{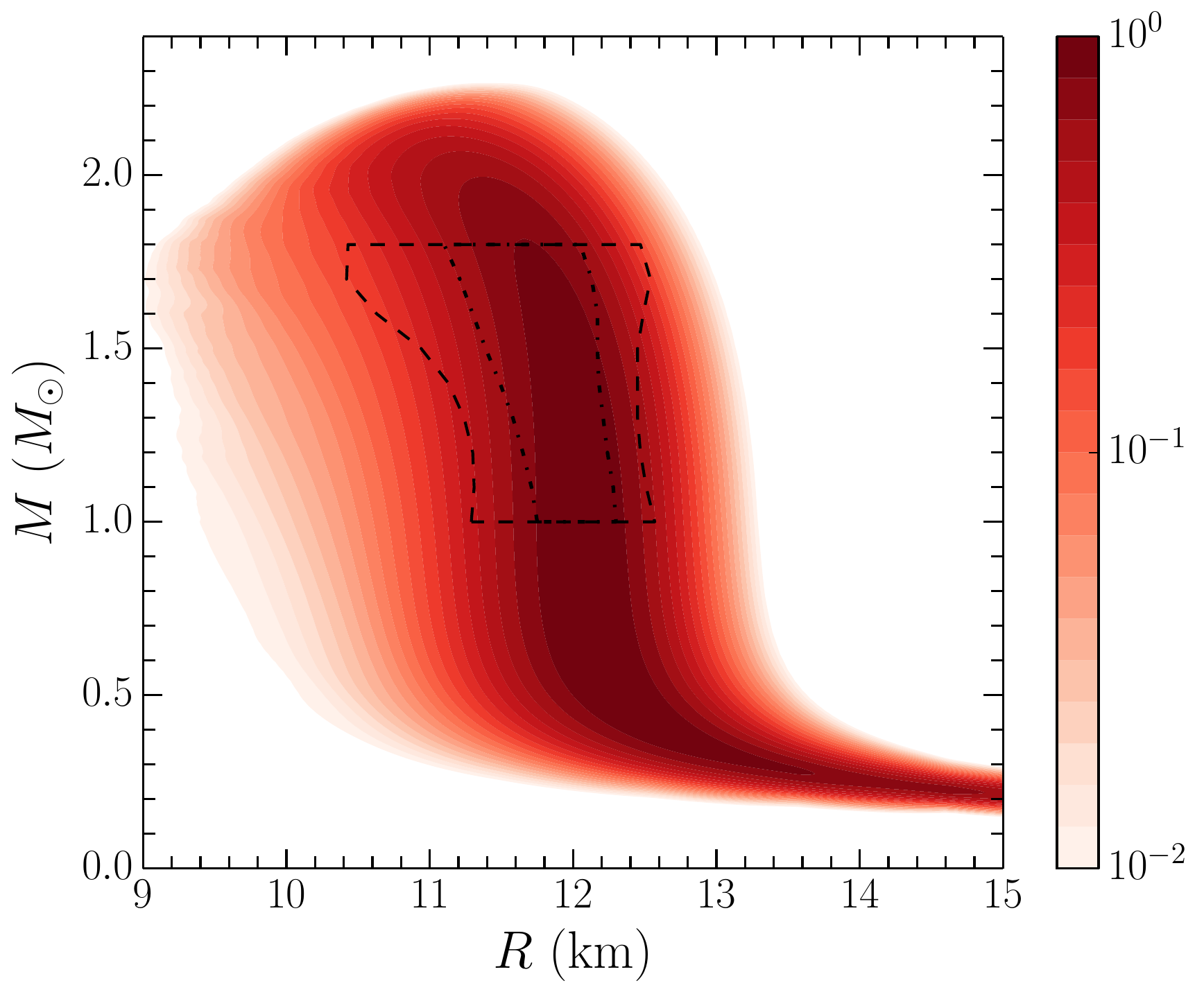}}\quad \quad
\subfigure{\includegraphics[scale=0.44]{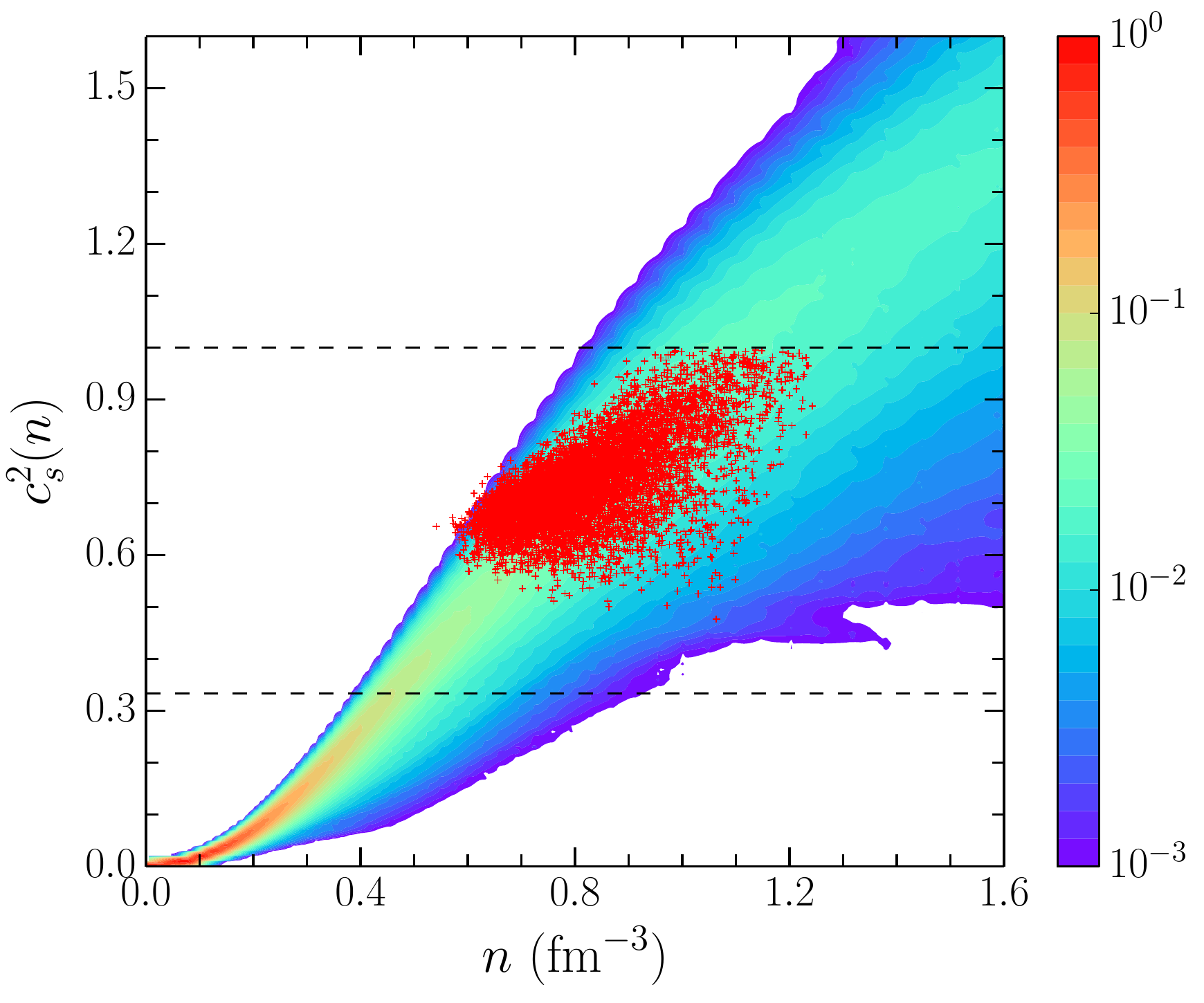}}
\caption{Left: Probability distribution for the neutron star mass-radius relationship obtained from the Bayesian posterior probability distribution for the nuclear equation of state. Right: Probability distribution for the speed of sound in neutron star matter as a function of the density. The red dots denote the central density and corresponding speed of sound for a $2M_\odot$ neutron star.}
\label{fig:mr}
\end{figure*}

In the right panels of Fig.\ \ref{prior}, we show the resulting probability distributions for the pure neutron matter (PNM) and symmetric nuclear matter (SNM) equations of state obtained from the posterior probability distributions for the $a_i$ and $b_i$ parameters. For both the neutron matter and symmetric matter equations of state, the prior distributions and likelihood functions are consistent, leading to reduced uncertainties in the posterior probability distributions. In particular, the properties of finite nuclei strongly constrain the symmetric nuclear matter energy per particle around saturation density $n_0$, which propagates to reduced uncertainties also at higher densities.

To construct the neutron star equation of state for both the crust and core under the condition of beta equilibrium, we interpolate between the pure neutron matter and symmetric nuclear matter equations of state using a quadratic approximation:
\begin{equation}
\bar{E}(n,x)= \bar E_{snm}(n) + S_2(n)(1-2x)^2.
\label{isodep}
\end{equation}
Generically, the energy per particle of isospin-asymmetric nuclear matter cannot be expanded in a Maclaurin series around symmetric nuclear matter ($\delta_{np} = (n_n-n_p)/(n_n+n_p) = 1-2x = 0$) due to the appearance of logarithmic contributions \cite{kaiser15,wellenhofer16} that give rise to a modified expansion
\begin{equation}
\bar{E}(n,x) = \bar{E}_{snm}(n) + S_2(n)(1-2x)^2 + \sum_{i=2}^{\infty}(S_{2i} + L_{2i} \ln \vert 1-2x \vert )(1-2x)^{2i}\,.
\label{isodep}
\end{equation}
Nevertheless, microscopic calculations have shown \cite{wellenhofer16,lagaris81,Bombaci91,drischler16b} that the contributions beyond $S_2$ are small and can be neglected. Constructing the inner crust equation of state is a phase coexistence problem in which a dense heavy nucleus is embedded in a dilute gas of unbound neutrons. We use the Wigner-Seitz cell formalism and treat the heavy nucleus (or nuclear pasta) within the liquid-drop model. The configuration of the system is then determined by minimizing the total energy with respect to the cell and heavy nucleus geometry. In particular, we construct a different model of the neutron star crust for each sampled equation of state. For a given equation of state we start from the crust when the total baryon number density is 10$^{-12}$\,fm$^{-3}$. When the neutron chemical potential is greater than zero, we then begin constructing the inner crust. When the energy density in the inner crust configuration is greater than that of uniform nuclear matter (or if the inner crust code does not converge), we find the equation of state from uniform nuclear matter. Additional details can be found in Ref.\ \cite{lim17}.

\section{Results}

\begin{figure}[t]
\centering
\includegraphics[scale=0.6]{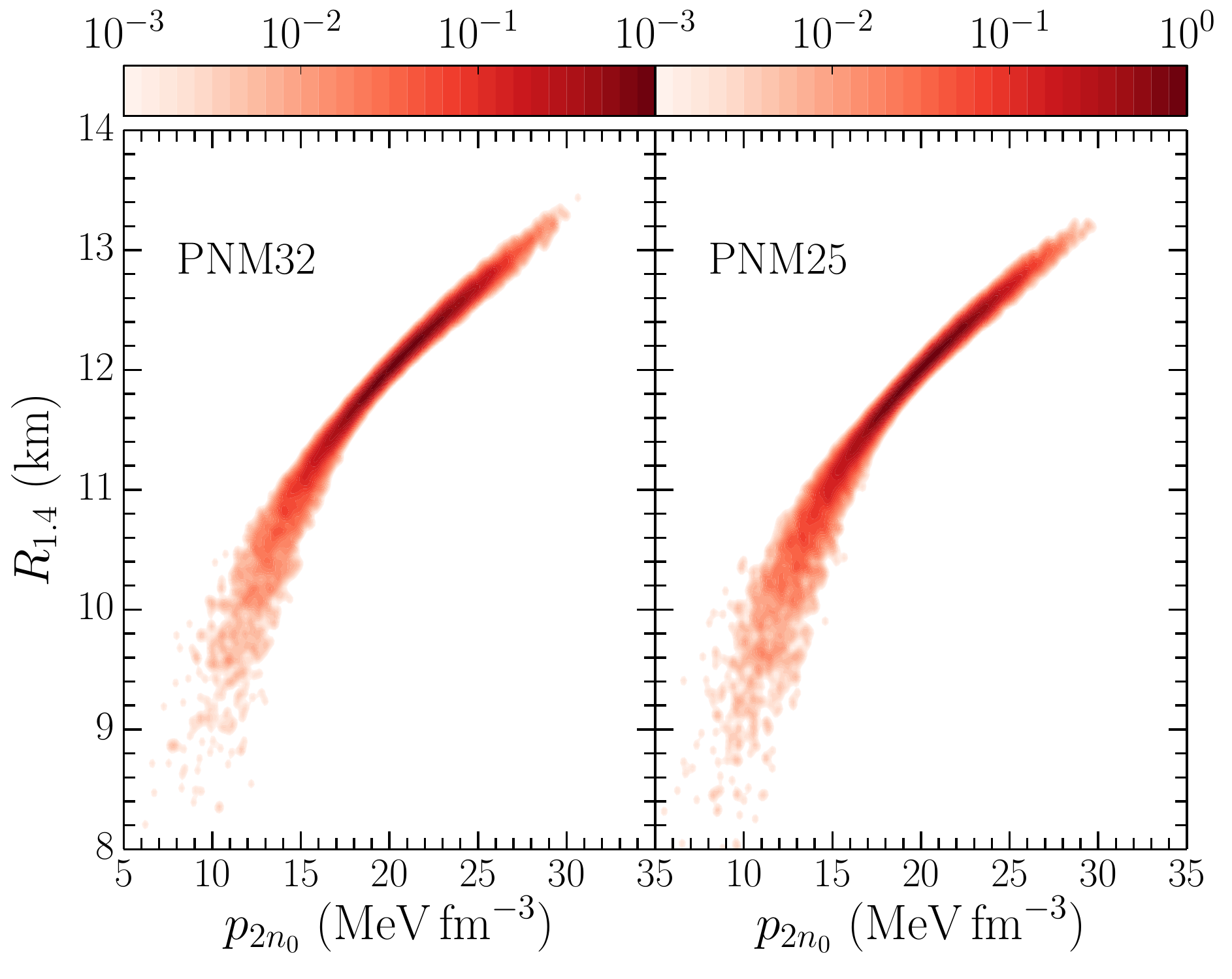}
\caption{Probability distribution for the radius of a $1.4\,M_\odot$ neutron star vs.\ the pressure at twice saturation density ($n=2n_0$) obtained from the Bayesian posterior probability distribution for the nuclear equation of state. Left: Prior probability distribution obtained by fitting chiral EFT equations of state up to $n=0.32$\,fm$^{-3}$. Right: Prior probability distribution obtained by fitting chiral EFT equations of state up to $n=0.25$\,fm$^{-3}$.}
\label{fig:rad}
\end{figure}

In the left plot of Fig.\ \ref{fig:mr} we show the resulting mass-radius relationship for 15,000 equations of state sampled from the Bayesian posterior probability distribution that includes constraints from nuclear theory and experiment. Equations of state that give rise to superluminal speeds of sound in the interior of neutron stars have been omitted. This applies to all stable neutron star configurations up to the maximum mass allowed for a given equation of state. For a $1.4\,M_\odot$ neutron star, the radius is found to lie in the range $10.27\,\rm{km} < R_{1.4} < 12.84\,\rm{km}$ at 95\% credibility. Although the resulting equations of state are relatively soft, in most cases ($\sim 75\%$) they can support $2M_\odot$ neutron stars. As seen in the left plot of Fig.\ \ref{fig:mr}, the probability to produce neutron stars heavier than $2.3M_\odot$ is very small. However, we note that the pressure support at very high densities $n>3n_0$ is poorly constrained in our modeling and therefore the inferred maximum masses in Fig.\ \ref{fig:mr} should not be interpreted too strongly. In particular, by construction there are no powers of the Fermi momentum beyond $k_F^6$ in the energy per particle and there are no quark-hadron phase transitions, hyperons, or other exotic states at high density. The present modeling therefore provides a baseline for comparison to scenarios containing novel phases of matter.

The probability distribution for the speed of sound as a function of density in beta-equilibrium matter is shown in the right panel of Fig.\ \ref{fig:mr}. We see that most of the equations of state produce superluminal speeds of sound at very high density, however such densities are unphysical and lie beyond the threshold for collapse to a black hole in all equations of state kept for statistical analysis. In the right plot of Fig.\ \ref{fig:mr} the red dots denote the central density and corresponding speed of sound for $2M_\odot$ neutron stars. This is merely a convenient choice of reference mass, and even more massive stable neutron star configurations have subluminal speeds of sound at all densities. The upper and lower dashed horizontal lines in the plot denote the speed of light ($c_s = c$) and the conformal limit value for the speed of sound ($c_s = c/\sqrt{3}$). As noted in Ref.\ \cite{bedaque15}, it is very difficult for the equation of state to support $2M_\odot$ neutron stars while observing a strict conformal bound $c_s(n) < c/\sqrt{3}$ for all densities $n$.

In the left plot of Fig.\ \ref{fig:rad} we show the probability distribution for the radius of a $1.4\,M_\odot$ neutron star as a function of the pressure of beta-equilibrium matter at $n=2n_0$ when the equation of state from chiral effective field theory is fitted up to $n=0.32$\,fm$^{-3}$. In the right plot we show the same probability distribution but when the equation of state from chiral effective field theory is fitted up to $n=0.25$\,fm$^{-3}$. It is somewhat surprising that the choice of maximum density does not strongly affect the final probability distributions. This is due to the smoothness of the chiral EFT equations of state and the fact that we simply extrapolate them to higher densities beyond the maximum of the fitting region. From Fig.\ \ref{fig:rad} a small ($\sim \!10\,\rm{km}$) or large ($\sim \!13\,\rm{km}$) neutron star radius measurement at the 5\% precision level, such as that expected from NICER, could significantly constrain our modeling of the equation of state. Less restrictive would be radius measurements in the range $10.5\,\rm{km} < R_{1.4} < 12.5\,\rm{km}$. Due to the strong correlations between the neutron star radius and the pressure seen in Fig.\ \ref{fig:rad}, precise radius measurements have the potential to strongly constrain the dense matter equation of state in a regime where microscopic nuclear modeling is currently uncertain.

\begin{figure}[t]
\centering
\includegraphics[scale=0.6]{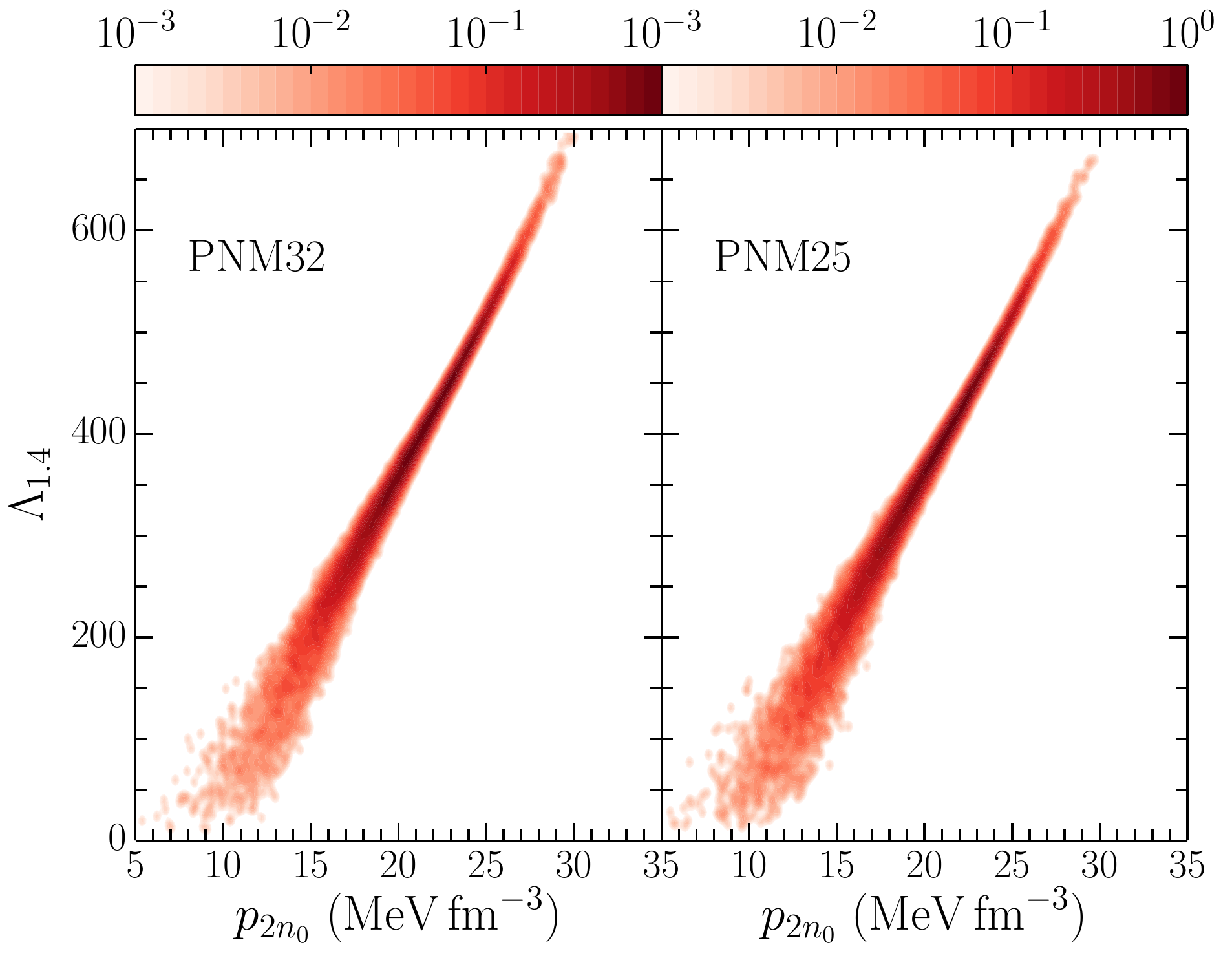}
\caption{Probability distribution for the tidal deformability $\Lambda$ of a $1.4\,M_\odot$ neutron star vs.\ the pressure at twice saturation density ($n=2n_0$) obtained from the Bayesian posterior probability distribution for the nuclear equation of state. Left: Prior probability distribution obtained by fitting chiral EFT equations of state up to $n=0.32$\,fm$^{-3}$. Right: Prior probability distribution obtained by fitting chiral EFT equations of state up to $n=0.25$\,fm$^{-3}$.}
\label{fig:tid}
\end{figure}

In the left plot of Fig.\ \ref{fig:tid} we show the probability distribution for the tidal deformability of a $1.4\,M_\odot$ neutron star as a function of the pressure of beta-equilibrium matter at $n=2n_0$ when the equation of state from chiral effective field theory is fitted up to $n=0.32$\,fm$^{-3}$. As in Fig.\ \ref{fig:rad} above, the right plot shows the same probability distribution but when the fitting range is taken up to $n=0.25$\,fm$^{-3}$. Again we find only a small difference in the results between the two fitting ranges. In both plots we see that the correlation at small pressures and small tidal deformabilities is somewhat weak, but for values of $\Lambda_{1.4} > 400$ there is a very tight correlation with the pressure at twice saturation density. In this region a 20\% uncertainty in the tidal deformability translates into a roughly 20\% uncertainty in the derived equation of state. At the 95\% credibility level, the tidal deformability lies in the range $100 < \Lambda_{1.4} < 500$, which is already consistent with 90\% credibility bounds $70 < \Lambda_{1.4} < 580$ from the re-analyzed  \cite{abbott2018b} GW170817 event.

\begin{figure}[t]
\centering
\includegraphics[scale=0.6]{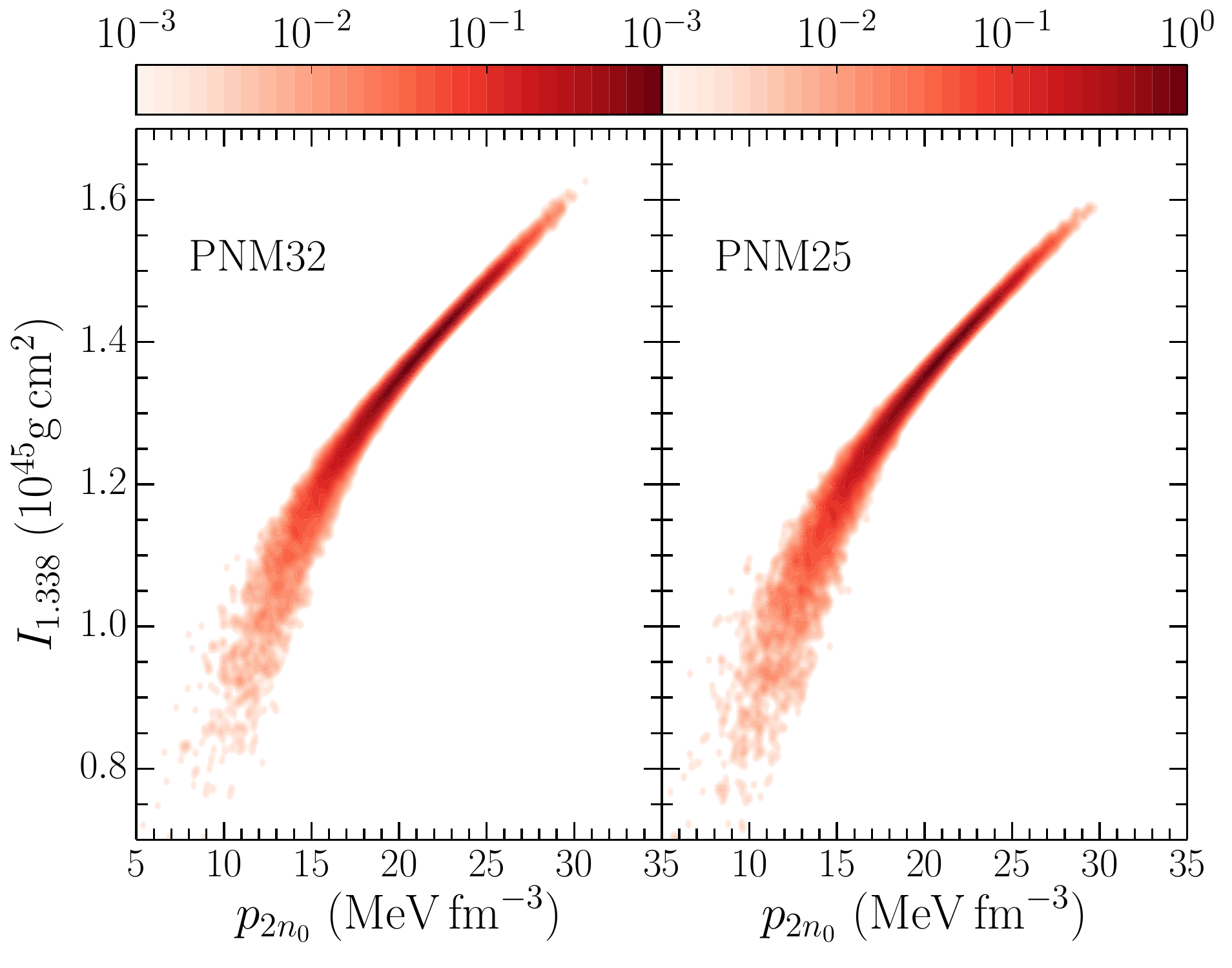}
\caption{Probability distribution for the moment of inertia of a $1.338\,M_\odot$ neutron star vs.\ the pressure at twice saturation density ($n=2n_0$) obtained from the Bayesian posterior probability distribution for the nuclear equation of state. Left: Prior probability distribution obtained by fitting chiral EFT equations of state up to $n=0.32$\,fm$^{-3}$. Right: Prior probability distribution obtained by fitting chiral EFT equations of state up to $n=0.25$\,fm$^{-3}$.}
\label{fig:moi}
\end{figure}

Finally, in the left plot of Fig.\ \ref{fig:moi} we show the probability distribution for the moment of inertia $I$ of a $1.338\,M_\odot$ neutron star as a function of the pressure of beta-equilibrium matter at $n=2n_0$. The right plot shows the same correlation but over the smaller fitting range $n < 0.25$\,fm$^{-3}$ for the chiral equations of state. The mass of $1.338\,M_\odot$ corresponds to the more rapidly rotating neutron star in the binary pulsar system PSR J0737-3039. It has been suggested \cite{lyne04,lattimer05} that radio timing observations of this pulsar could lead to a precise measurement of periastron advance. Relativistic spin-orbit coupling, which depends on the neutron star moment of inertia $I$, contributes a small amount to the periastron advance, entering at second order in a post-Newtonian expansion of the orbital motion. Constraints on the neutron star moment of inertia at the 10\% precision level therefore may be possible from high precision radio timing. In our theoretical modeling, we find that at the 95\% credible level the moment of inertia lies in the range $1.03 \times 10^{45}\,\rm{g cm}^2 < I_{1.338} < 1.51 \times 10^{45}\,\rm{g cm}^2$. From Fig.\ \ref{fig:moi} we see that there is a very tight correlation between the moment of inertia and the pressure of beta-equilibrium matter at $2n_0$, especially for larger values of $I$. An astrophysical measurement of $I_{1.338}$ could therefore provide a novel consistency check complementary to neutron star radius measurements from NICER and tidal deformabilities from the LIGO/Virgo collaboration.

\section{Conclusions}
We have shown how Bayesian modeling of the dense matter equation of state incorporating constraints from nuclear theory and experiment allows for a statistical analysis of neutron star radii, tidal deformabilities, and moments of inertia. All three of these quantities are under investigation with present observational campaigns of neutron stars. We have demonstrated strong correlations among these bulk neutron star properties and the pressure of neutron-rich matter at twice saturation density. The developed statistical framework will therefore enable constraints on the nuclear matter equation of state in a novel density regime from future observations of neutron stars.

\section{Acknowledgement}
Work supported by the National Science Foundation under grant No.\ PHY1652199. Portions of this research were conducted with
the advanced computing resources provided by Texas A\&M
High Performance Research Computing.



\end{document}